\documentstyle[11pt,aas2pp4]{article}
\newcommand{\vsc} {$r_{\rm inj}\,\Omega_{\rm inj}$}

\newcommand{\lsi} {$r_{\rm inj}$}
\newcommand{\tsc} {$\Omega_{\rm inj}^{-1}$}
\newcommand{\bvec}[1] {{\bf {#1}}}
\lefthead{J.R. Murray, M. de Kool \& J. Li}
\righthead{Counter-rotating disks and spin-up spin-down}
\begin{document}
\title{Accretion disk reversal and the spin-up/spin-down of accreting pulsars}	
\author{James R. Murray, Martijn de Kool and
Jianke Li}
\affil{ Astrophysical Theory Centre\footnote{Operated jointly by Mount
Stromlo and Siding Spring Observatories \& School of Mathematical
Sciences, The Australian National University, ACT 0200, Australia}, 
Australian National
University, ACT 0200, Australia}	
\begin{abstract}
We numerically investigate the hydrodynamics of accretion disk
reversal and relate our findings to the observed spin-rate changes in
the accreting X-ray pulsar GX~1+4. In this system, which accretes from
a slow wind, the accretion disk contains two  dynamically distinct regions.
In the inner part viscous forces are dominant and disk
evolution occurs on a viscous timescale. In the outer part
dynamical mixing of material with opposite angular momentum is
more important, and the externally imposed angular momentum reversal 
timescale governs the flow. In this outer region the disk is split into 
concentric rings of
material with opposite senses of rotation that do not mix completely but
instead remain distinct, with a clear gap between them. 
We thus predict that torque reversals resulting from
accretion disk reversals will be accompanied by minima in accretion
luminosity. 
\end{abstract}
\keywords{
          accretion, accretion disks --- binaries: general --- hydrodynamics
          --- methods: numerical --- 
          pulsars: individual (GX~1+4) --- stars: rotation}
\section{Introduction}

The availability of continuous monitoring data for several 
X-ray pulsars as obtained with the BATSE All Sky Monitor on the
Compton Gamma Ray Observatory over the last five years
(Bildsten et al. 1997, Nelson et al. 1997) has led to renewed interest
in the spin evolution of accreting neutron stars 
with a strong magnetic field. From earlier pointed observations of
X-ray pulsars with other satellites it was already known that the 
spin rates of these neutron stars show changes on many timescales, 
ranging from unresolved fluctuations
(shorter than 1 day) to systematic trends lasting over decades (see, e.g.
the compilation by Nagase 1989). 

X-ray pulsars are usually divided into two groups, those which accrete
from a Roche-lobe filling companion by way of an accretion disk, and
those that accrete from the stellar wind of a companion star, in which
case a disk may or may not be present. The second group tends to show
more erratic spin rate changes on longer timescales, whereas the disk
accreting pulsars generally show long term systematic trends. On the
shortest timescales however the spin rate changes in the two groups
appear to be comparable (de Kool \& Anzer 1993, Baykal \& \"Ogelman 1993).

Several models have been proposed to explain these spin rate
fluctuations, or alternatively the torques that cause them. In the
standard disk accretion model (Ghosh \& Lamb 1979) the pulsars are
spinning at an equilibrium period, at which the spin-up torque due to 
the angular momentum of the accreting material (often referred to as
the material torque) and the interaction of the magnetic field of the
pulsar with the disk inside the co-rotation radius is balanced by a
braking (spin-down) torque due to the interaction  with the more
slowly rotating outer parts of the
accretion disk. The equilibrium period depends on the accretion rate,
and the pulsar is expected to spin-up or down as the accretion rate 
increases or decreases. This model appears consistent with the observed
positions of X-ray pulsars in the pulse period - X-ray luminosity
(equivalent to accretion rate) diagram. It also explains why the long
term spin rate changes are considerably smaller than what would be
expected from the material torque alone.

However, the standard disk accretion model does not explain
the spin rate fluctuations observed on very short timescales that are
superposed on the long timescale trends in most disk accreting
sources. These fluctuations must be caused by torques that are as
strong as the full material torque, much larger than the net torques
expected to result from a small disequilibrium between spin-up and
spin-down torques. 

For the wind-accreting pulsars, a different mechanism can be invoked
to explain the positive and negative spin rate changes. Early numerical
studies of Bondi-Hoyle accretion flow (e.g. the two dimensional
simulations  by Matsuda  et al. 1987, Fryxell \& Taam 1988, 1989) 
demonstrated that it may be possible 
to form temporary accretion disks with alternating senses of rotation 
in a wind accreting system. More recent high resolution
three dimensional numerical investigations of the so-called wind
``flip-flop'' instability may be found in Ruffert (1992, 1997) and the
references therein. The timescale of the disk reversals is found to be
of the order of the sound crossing time of the accretion radius, which for
the most common systems in which the neutron star accretes from a fast
wind from an early type (O or Be) star is of the order of hours. This is
consistent with the torque fluctuations observed in wind accreting
sources, which also appear to be of the order of the material torque
discussed above, but applied in random directions with the direction
changing on a timescale shorter than 1 day.

The new continuous monitoring data obtained by BATSE 
(Bildsten et al. 1997, Nelson et al. 1997, Chakrabarty et al. 1997)
have yielded values for the mass accretion rate (X-ray
luminosity) and the accretion torque (change in spin frequency) 
on a regular basis for a large number of X-ray pulsars.
This data set has allowed a detailed comparison of the observed relation
between accretion rate and torque, and that predicted by theoretical
models. (It should always be kept in mind however that the fluxes measured by
BATSE are only the {\em pulsed} fraction of the X-ray luminosity, and that
it is not clear whether the ratio between pulsed and non-pulsed fluxes
is constant).

The observations are not consistent with the standard Ghosh \&
Lamb model, since this predicts a clear correlation between spin-up
and an increase in X-ray luminosity, whereas the observations show a
variety of behaviour, with spin-up or spin-down occurring at the same
apparent luminosity. The cause of this type of behaviour in disk accreting
systems is not well understood, but a possible explanation has
recently been suggested by some of us (Li \& Wickramasinghe 1998).  

However, for wind accreting systems the possibility of accretion
from temporary accretion disks with alternating senses of rotation
is still tenable. In analogy,
Nelson et al. (1997) made the {\em ad hoc} suggestion that many
observational features of some systems that are normally thought to
contain disks (GX~1+4, 4U~1626-67) would be explained if they were 
accreting from disks with alternately prograde and retrograde senses of
rotation. Previously, Makishima et al. (1988), Dotani et al. (1989)
and Greenhill et al. (1989)
had also sought to explain the rapid spin-down of GX~1+4 in terms of
accretion from a retrograde disk. 

If the secondary star is feeding the
accretion disk via Roche lobe overflow, as is almost certainly the
case in 4U~1626-67, it is hard to conceive how a
retrograde disk could ever come about. However, in the case of GX~1+4,
the suggestion is not unreasonable. This X-ray pulsar is unique in the
sense that it is accreting from a red giant or AGB star wind (Chakrabarty
\& Roche 1997), and is in a very wide orbit. Estimating the timescale
of disk reversal for accretion from such a wind, one obtains a
timescale of the order of years, and the disk would form at a large radius
($\sim 10^{13}$ cm) so that the inner part of the accretion flow is
expected to be like a normal accretion disk. A timescale of years
corresponds well with the timescale on which the accretion
behaviour in GX~1+4 is observed to change, with a negative correlation
between accretion rate and spin-up in some phases while the disk would
be retrograde, and a positive one at other times when it is prograde
(Chakrabarty et al. 1997). Thus, this system is ideally suited to
study the possibility of forming retrograde disks, since the timescale
for disk reversal would be much longer than that of 
the torque fluctuations on a timescale of one day or less that are
common in all types of X-ray pulsars. 
%In our opinion, these are more
%likely to be due to some instability in the interaction between
%accretion disk and magnetosphere. 
In the systems that accrete from a
fast wind, the two timescales are comparable, and the effects will
be difficult to separate.  

Relatively little modelling of the hydrodynamics of disk reversal
has been performed, making it difficult to assess whether this model 
is in agreement with the observations. The previous numerical 
studies of Bondi-Hoyle accretion flow (e.g. Ruffert 1997, Matsuda et 
al. 1987)
did not resolve the inner part of the accretion disk, and did not
yield much insight in such details as the interaction of an existing disk
with material coming in with opposite angular momentum, the timescales
determining the evolution of such systems, and the relation
between accretion torque and luminosity expected in such a model.

In this paper we use Smoothed Particle Hydrodynamics (SPH) 
simulations to start an investigation
into these processes, and compare our results with the observation of
GX~1+4.

\section{Numerical Results}
\subsection{Numerical Method}
The aim of the calculations described here is to investigate the
response of an accretion disk to a change in the specific angular
momentum of the material it is being fed. 
In particular we are interested in the case were the specific 
angular momentum of the material added at the disk's outer boundary is
reversed. 

The simulations were performed using an
SPH code that has been specifically
adapted for disk problems (Murray 1996).
Two and three dimensional
applications of the code can be found in Murray  (1998) and
Murray \& Armitage (1998). To control the computational expense,
motion perpendicular to the disk plane was neglected, and the
calculations were run in two dimensions. We took full advantage of the
adaptive nature of SPH and allowed the particle smoothing length (and
hence the resolution) to vary in space and time.

Viscous dissipation is included in the code using a linear term in the
SPH equations that is based upon the linear artificial viscosity term
described in Monaghan (1992). Detailed descriptions and tests of the term's
application to accretion disks are to be found in Murray (1996,
1998). In the continuum limit this dissipation term introduces a
viscous force per unit mass 
\begin{equation}
\bvec{a}_{\rm v}=\frac{\zeta h }{8\Sigma}\,
(\bvec{\nabla}\cdot(c \Sigma \bvec{S})+\bvec{\nabla}
(c \Sigma \bvec{\nabla} \cdot \bvec{v})),
\label{eq:genvisc}
\end{equation}
where $\bvec{S}$ is the deformation tensor,
$\Sigma$ is the column density, $h$ is the smoothing length, $c$ is
the sound speed, and $\zeta$ is the SPH linear artificial viscosity
parameter. In this work we have simply set $\zeta=1$. In the
simulations described below, both the density and the sound speed
vary on similar spatial scales to the velocity. Hence we cannot (as
was previously done in e.g. Murray 1998) approximate the dissipation in
equation~\ref{eq:genvisc} as the sum of a shear and a bulk viscosity.
\subsection{Simulation Details} 

We simulated the gas flow in the annular region $r_{\rm in} <
r < r_{\rm inj}$. $r_{\rm inj}$ is the radius at which
gas is added to the simulation. Ideally we should like $r_{\rm in}$ to
be of the order of the magnetospheric radius $r_{\rm m}$ (a radius
characterising the truncation of the viscous accretion disc), 
and $r_{\rm inj}$
to be of the order of 0.1  times the accretion radius, $$r_{\rm
 acc}=G\,M/v_{\rm w}^2,$$ where $G$ is the gravitational constant, $M$ is
the mass of the neutron star, and $v_{\rm w}$ is the wind velocity.
Indeed for sources accreting from a fast
wind the disk formation radius is not much larger than $r_{\rm m}$ and it
is computationally feasible to simulate the entire region 
$r_{\rm m} < r < 0.1 \,r_{\rm acc}$. 

However, for the system GX~1+4, the accretion
radius is several orders of magnitude larger than $r_{\rm m}$, and we
can not resolve the full range in radius over which the disk is likely
to extend. We have worked around this problem in the following way.
As discussed in the next section, we expect the disk to be divided in
two parts, an inner one in which viscous forces dominate, and an
outer one in which dynamical mixing of material with opposite angular
momentum is more important. We therefore consider two cases, one with
a very low viscosity which should apply to the outer regions of the
disk, and another with much higher viscosity which models the
parts of the disk where the transition from dynamical mixing to
viscous transport occurs. Although this still does not cover the range
in radius all the way down to the magnetospheric radius, we can expect
that inside the transition region a standard accretion disk is present
which adjusts instantaneously to its outer boundary condition.

The gravitational potential was simply taken to be that of the
accreting star, thereby neglecting tidal effects. 
This is likely  a very good approximation, since the hydrodynamical
simulations of the flip-flop instability show that disk formation
generally occurs at $r \sim 0.1 \,r_{\rm acc}$, and that $r_{\rm acc}$ is
always smaller than (or at most comparable to) the Roche lobe radius.

We chose a cylindrically symmetric mass inflow condition at the outer boundary.
Single particles were added to the calculation at ten points, 
evenly spaced around a circle of radius $r_{\rm inj}$ centred on the neutron
star. Each particle was set moving in the azimuthal direction with
specific angular momentum $j$, corresponding to that of a Kepler orbit
of radius $r_{\rm circ}$.
 Particles were added in this fashion at 
regular time intervals
$\Delta t$. 

We did not attempt to model 
the details of the
disk-magnetosphere interaction, and simply  used a free inner boundary
at $r=r_{\rm in}$. At the end of each time step, any particle at a radius 
$r< r_{\rm in}$ was assumed to have been accreted by the central object. 
The mass and angular momentum of the particle was recorded, and
the particle was removed from the calculation. 
For computational reasons, particles
at very large radii $r > r_{\rm out}$ were removed in a similar
fashion. 
However, in this calculation the mass (and angular momentum) lost at
large radii was too small (less than 1\% of the total angular
momentum) to significantly effect the evolution of the disk.

The temperature profile of a steady state disk was imposed upon the
flow. Thus, the sound speed
\begin{equation}
c=c_0\,(\frac{r}{r_{\rm circ}})^{-3/8}
\end{equation}
where $c_0$ is the sound speed at the circularisation radius. We found
that details of the disk's temperature profile did not significantly
affect the results described below. In other words, pressure
effects were not important.

In the simulation we have scaled quantities so that the mass of the
accreting neutron star, $M$, the radius of mass addition, 
$r_{\rm inj}$, and the Keplerian angular velocity at that radius,
$\Omega_{\rm inj}$, are unity.
The values of the various parameters  used in this calculation are
given in Table~\ref{tbl:params} (in terms of our scaled units).

%\subsection{Results}

\subsection{Evolution with non-negligible viscosity}

We obtained the initial condition for this calculation by running the
code for $200$ \tsc\, with mass added in an anti-clockwise
direction as described above. At the end of this initialisation (time $t=0$) 
the disk contained 16694 particles. 

The calculation proper was started by altering the mass addition
routine so that newly added material flowed in the clockwise direction
(with the same magnitude $j$ as before).
 The simulation was then run for a further $200$ \tsc.

\begin{figure}
\plotone{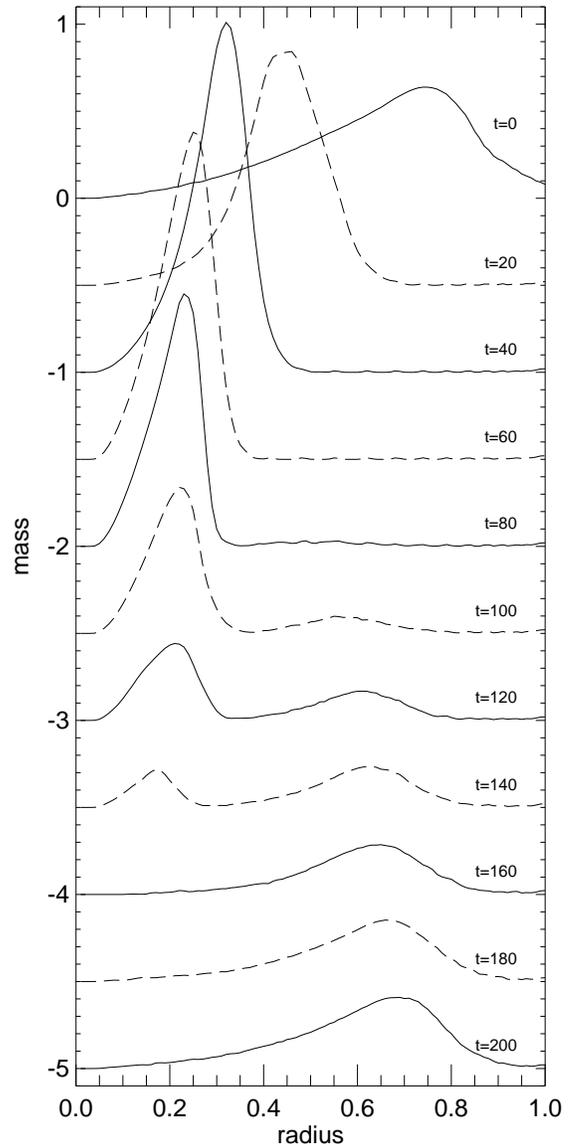}
\caption{Mass as a function of radius for the high viscosity
calculation (section 2.2), at the times shown. 
The angular velocity of mass added at the outer boundary is
reversed at $t=0$ (topmost curve). Subsequent curves are offset 0.5
units in mass.
Every second curve is dashed for clarity.}
\label{fig:mprofiles}
\end{figure}

\begin{figure}
\plotone{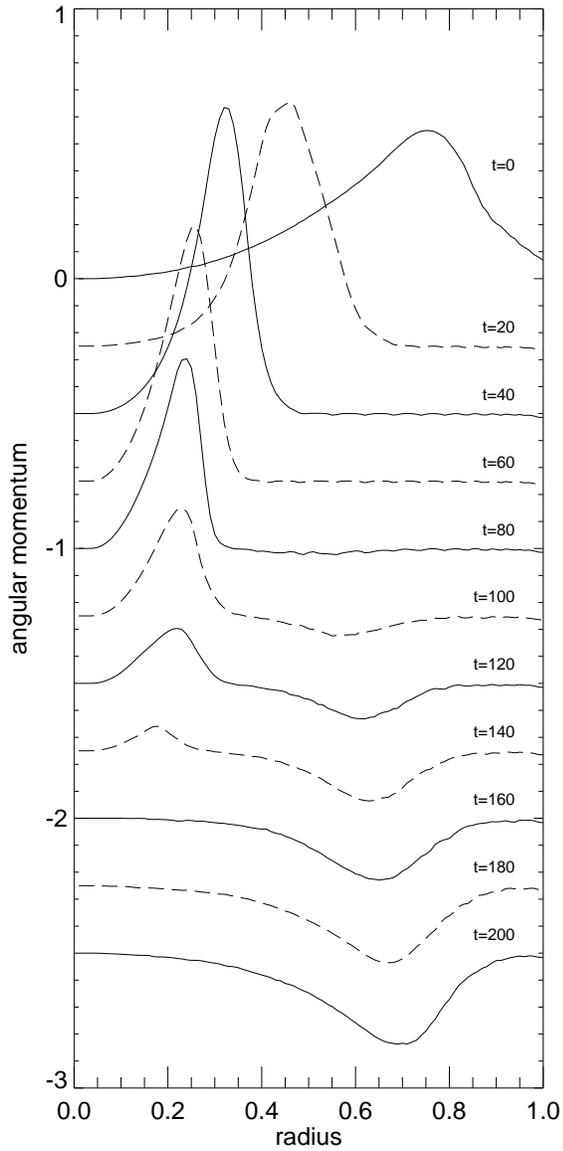}
\caption{Angular momentum as a function of radius for the high
viscosity calculation. 
Every second curve is dashed for clarity.}
\label{fig:amprofiles}
\end{figure}

\begin{figure}
\plotone{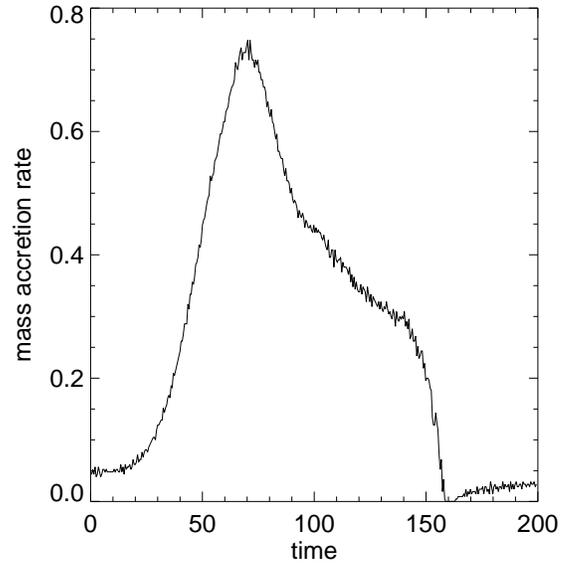}
\caption{Mass accretion rate as a function of time for the high
viscosity calculation.}
\label{fig:mass}
\end{figure}

\begin{figure}
\plotone{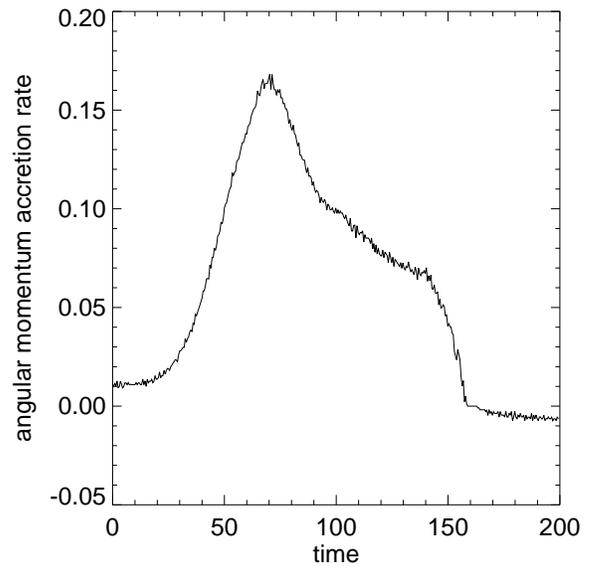}
\caption{Angular momentum accretion rate as a function of time for the
high viscosity calculation.}
\label{fig:am}
\end{figure}

The evolution of the disk's radial mass and angular momentum profiles from the
commencement of clockwise mass addition are shown in
Figs~\ref{fig:mprofiles} and~\ref{fig:amprofiles}
respectively. The  rates of mass and angular momentum
accretion  onto the central object are shown as functions of time in
figures~\ref{fig:mass} and~\ref{fig:am}. 
For approximately the first $70$ \tsc\, of the calculation, the
injection streams {\em directly impacted} the disk, compressing it inwards.
Fig~\ref{fig:mprofiles} clearly shows the radial mass distribution
becoming more strongly peaked  over time, with the radius of maximum mass
moving rapidly inwards. Correspondingly the mass accretion onto the
central object increased to a maximum at time $t\simeq70$ \tsc. 
This initial disk
compression occurred on a timescale shorter than the outer disk's
viscous timescale. The collision of the injection streams with the
disk enforced a very efficient exchange of angular momentum between the
two flows. The specific angular momentum of the gas at  the outer edge
of the disk
was thus sharply reduced. This material then moved to smaller radii,
with radial pressure gradients and internal viscous torques too small
to prevent the disk shrinking. 

Eventually, the disk was compressed to a size where the injection
streams could no longer impact directly upon it. Consequently the disk
shrank no further and the mass accretion dropped sharply away.

It is easy to show that  a
ballistic particle, launched in the azimuthal direction at radius
$r_{\rm inj} $ with the specific angular momentum of a Kepler orbit of
radius $r_{\rm circ}$, will reach periastron at radius
\begin{equation}
r_{\rm p}=\frac{r_{\rm inj}\, r_{\rm circ}}{2r_{\rm inj}- r_{\rm circ}}.
\end{equation}
For our calculation, $r_{\rm p} = r_{\rm inj}/3$. Indeed, 
Fig~\ref{fig:mprofiles} shows that the disk is compressed to
this radius at around $t \simeq 70$ \tsc. 
The minimum compression timescale is simply
obtained by ignoring internal stresses and calculating the angular
momentum that must be added to reduce the outer disk (plus newly added
gas) to a circular orbit of radius $r_{\rm p}$. For this calculation $\Delta
t_{\rm min} \simeq 30$ \tsc. In fact by $t=60$ \tsc\, all particles in
the calculation
remaining at radii $r > r_{\rm inj}/3$ already had their specific angular
momentum reduced below that of Kepler orbit with radius
$ r_{\rm inj}/3$.

Meanwhile, the accretion streams were no longer hindered by
the original disk, and were free to form a second disk. Initially a
narrow ring formed at $r=r_{\rm circ}$. Both the outer and inner ring
started to spread viscously, and the inner ring began to slowly
accrete through $r_{\rm in}$. Once the initially almost empty gap
between the two rings filled via viscous diffusion, the 
very strong shear between the two disks allowed angular momentum to be 
more efficiently extracted from the inner disk, and accelerated its
demise. Hence the final drop in mass
accretion was sudden rather than a gradual tailing away.

The angular momentum accretion rate in this simulation was simply
proportional to the mass accretion rate as is to be expected if gas
at $r_{\rm in}$ is still in Keplerian motion. Not until the inner disk
is entirely consumed does the accretion torque on the central object
change sign.

%\begin{figure}
%\plotone{dens2.eps}
%\caption{The radial density profile of the disk at the times shown.
%The angular velocity of mass added at the outer boundary is
%reversed at $t=0$ (topmost curve). Subsequent curves are offset one
%unit in density. Every second curve is dashed for clarity.}
%\label{fig:dprofiles}
%\end{figure}

%\subsubsection{Evolution with negligible viscosity}
\subsection{Evolution with negligible viscosity}

If viscous interaction in the disk is not important, the evolution of
the disk must be different from the one discussed above, in which the
behaviour during the
second half of the time interval modelled was dominated by viscous
effects. We therefore performed further simulations with a much lower
viscosity, obtained by reducing the sound speed by a factor of 5. The
maximum smoothing length, $h_{\rm max}=0.01\,r_{\rm inj}$. As a result
(see  equation~\ref{eq:genvisc}) the viscous forces are approximately a
factor ten smaller than in the previous simulation.

To save computation time, we used insight gained from the previous 
calculation and began with two, well separated, oppositely rotating rings with
identical Gaussian density profiles centred at $r_{\rm circ}$ 
and $0.5\, r_{\rm circ}$. New mass was then injected at the
outer radius with a sense of rotation opposite to that of the outer
ring, to simulate a second reversal of rotation at the outer
boundary. Mass resolution was improved by increasing the rate of
particle injection by a factor 5 over the first simulation. 

In fact we
completed three separate simulations for this section in order to
verify that the regions of interest were being adequately
resolved, and that are results were not sensitive to the numerics. The
first calculation began with  16826 particles in the two rings
combined. As in the previous section, particles were added at regular
intervals from ten equally spaced points around a unit circle, with
the points being fixed in the inertial frame. A second calculation was
completed
with identical system parameters but four times the mass resolution
(the initial set up contained 57680 particles). The rate of {\bf mass
addition} was  identical to the first simulation. 
A third calculation was done in which the
mass resolution was identical to the first calculation but the mode of
mass addition was varied. Particles were again added at ten equally
spaced points around the unit circle, but each set of ten was randomly
offset in azimuth. 
This was done to confirm that the instability
described below was real and not forced by the periodic nature of the
stream impact points on the outer ring. 
We only identified minor differences
in the results of the three simulations. 

\begin{figure}
\plotone{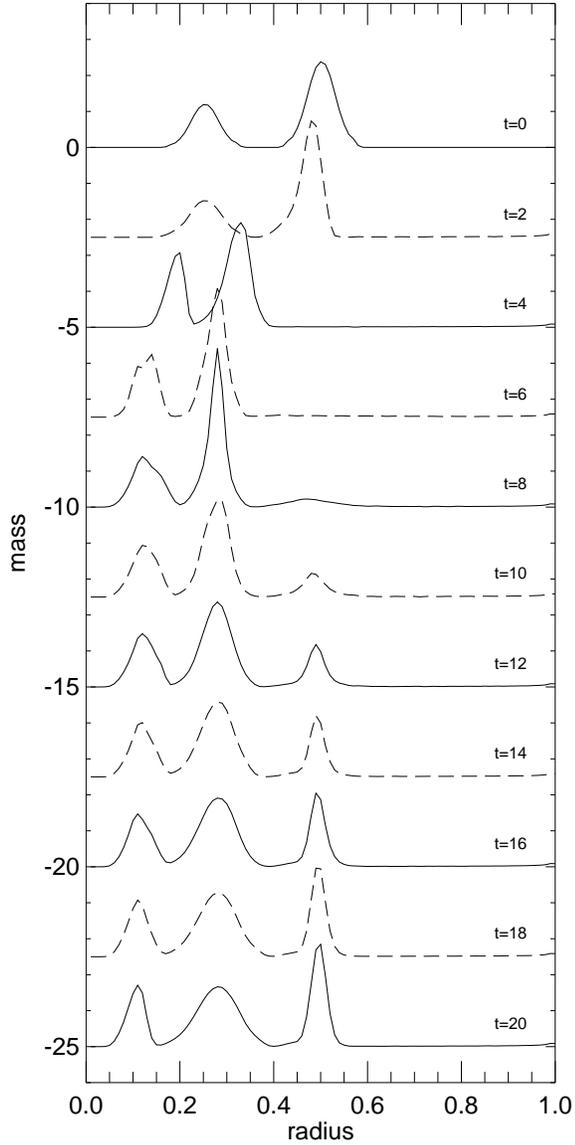}
\caption{Evolution of the radial mass profile for the third calculation with
negligible viscosity (section 2.3) in which mass injection was ``disordered'' 
(i.e. the azimuth of each set of
ten particles was set randomly). 
The calculation begins ($t=0$) with two counter-rotating 
Keplerian annuli with identical Gaussian density profiles centred at
$r=0.25$ and $r=0.5\,r_{\rm out}$.}
\label{fig:mm}
\end{figure}

\begin{figure}
\plotone{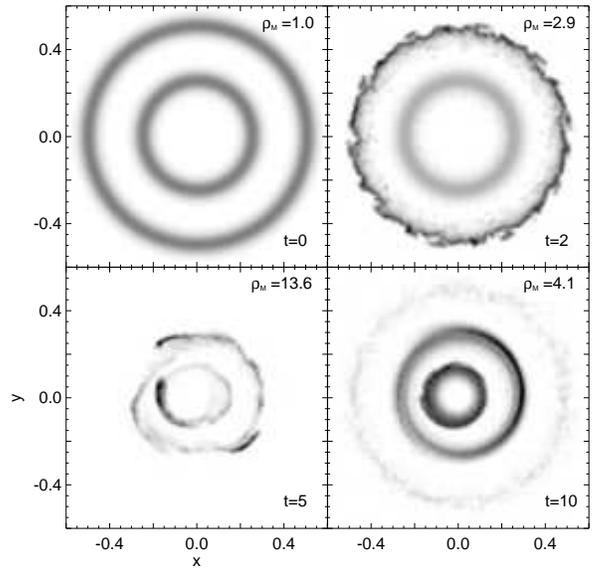}
\caption{Density maps of the ``disordered injection'' negligible viscosity
calculation shown in figure~\ref{fig:mm}. In order to show as much detail in
each frame, the maps have been separately scaled.
The maximum density and time of each snapshot are shown in the top
and bottom right respectively of each panel.}
\label{fig:dens4}
\end{figure}

\begin{figure}
\plotone{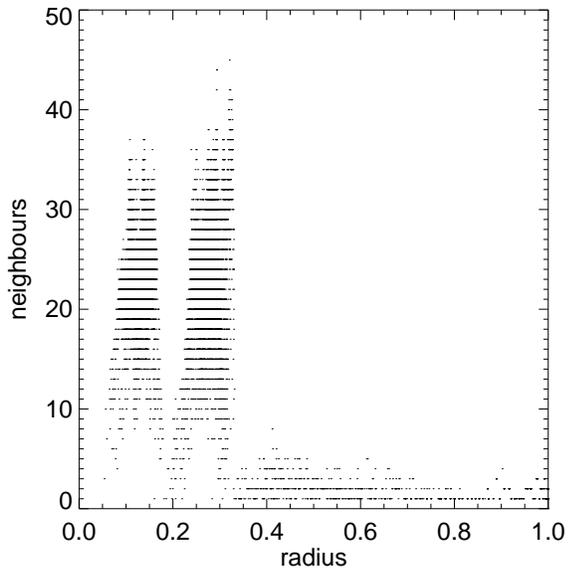}
\caption{Neighbour number plotted against radius for each particle at
$t=3$~\tsc\, in
the negligible viscosity simulation
with ``disordered'' mass injection.}
\label{fig:gapneigh}
\end{figure}

The evolution of the mass distribution for the third calculation
is shown in figure~\ref{fig:mm}.
As in the simulation described in the previous section, 
the outer ring moved in on the mixing timescale,
but in this case it encountered the innermost ring. 
For this low-viscosity
case, we found that the strong shear between the rings gave rise to a
non-axisymmetric instability (see figure~\ref{fig:dens4}) 
that deformed both rings severely. This 
instability mixed material with opposite angular momentum very
effectively, so that the inner ring also fell in. The severe
deformation was seen in all three simulations completed for this
section.
Once the outer ring 
reached the radius $r_{\rm p}$ and was not forced in any longer, an empty 
gap between the rings formed, the rings rapidly became circular
again, and a new ring began forming at the circularisation
radius. In figure~\ref{fig:gapneigh} we have plotted particle
neighbour number against radius for the third calculation at time
$t=3$ \tsc\, (by which stage the rings have largely recircularised and the
gap between them has formed). At this time only 260 of the 16864
particles at radii $r < 0.3$ \lsi\, have fewer than ten
neighbours. This, together with the fact that similar results were
obtained in the higher resolution simulation, indicates that the
calculations were not under-resolved.

We conclude from these calculations that if the externally
imposed mass reversal timescale is significantly shorter than the
viscous timescale at the circularisation radius, a number of
concentric rings with alternating senses of rotation could be present
inside the circularisation radius, down to the radius where the viscous
timescale for the formation of an accretion disk or the interaction 
between two peaks becomes shorter than the reversal timescale.

\section{Discussion}

\subsection{Timescales}
From our results it is clear that there are two processes governing
the inward transport of matter through the accretion disk. One is the
usual viscosity, and the other one is mixing of material with
different (or opposite) specific angular momentum. A first estimate of
the relative importance of these two processes in a physical system 
can be found by comparing the
timescales on which they can change the disk structure. 

The viscous timescale for a standard disk is given by 
\begin{equation}
\tau_v \sim {{r^2}\over{\alpha c_s h}} 
\sim 7 \times 10^6 \ \alpha_{0.1}^{-1} \  r_{10}^{3 \over 2} \ \left( h \over r
\right)^{-2}_{0.01} {\rm sec},
\end{equation}
in which $\alpha_{0.1}$ is the Shakura-Sunyaev viscosity parameter in
units of 0.1, $r_{10}$ the radius in units of $10^{10}$\ cm, and $h$
is the thickness of the disk. 
For a system like GX 1+4 this timescale is of the order of 7000 years
if we assume that the disk forms at a radius of about $0.1\,
r_{\rm acc} \sim 10^{13}$ cm. Note that counter-rotating rings will
interact viscously on a timescale shorter than that given by the
expression above, and can be estimated by replacing $h/r$ by $h/l$, where
$l$ is the separation between the two peaks

The typical timescale on which the inflow conditions at the 
outer boundary conditions can change due to the flip-flop instability
is 
\begin{equation}
t_{ff} \sim {{G M}\over {v_w^2 c_{s,w}}} \sim 9 \times 10^7
\ v_{w,15}^{-2}\  c_{s,10}^{-1} \ {\rm sec}
\label{eq:freefall}
\end{equation}
in which $v_{w,15}$ is the wind speeds in units of 15 km/s, and 
$c_{s,10}$ the sound speed in the wind in units of 10 km/s.
The timescale needed by the dynamical processes associated
with the change in boundary conditions to cause the disk
structure to change significantly must be the same as $t_{ff}$, since a
significant change requires mixing material already present in the
disk from the previous episode with a similar amount of material with
different angular momentum.

Again for the parameters of GX~1+4, we expect a steady viscous disk to
form when $t_v < t_{ff}$, implying $ r < 5 \times 10^{10}$\ cm if
$\alpha$ and $h/r$ have their typical values. For the pulsars
accreting from a fast wind this limit corresponds to $ r < 10^8$\
cm, which is smaller than or comparable to the magnetospheric radius
so that in these cases viscous forces must play a negligible role in
the accretion flow even if a disk forms.

These considerations led us to examine two cases in the numerical
simulations. In the first, the viscosity was so large that it governed
the evolution near the inner boundary of the computational domain,
which modelled the transition from the mixing dominated to the viscosity
dominated regime at small radii. The second simulation addressed what
happens in the outer parts of the disk where viscosity is unimportant.
In both cases we found that it is always the externally imposed
timescale that determined the timescale of changes at the inner
boundary. If the viscosity was low,
leading to a viscous timescale much longer than the external reversal
time, dynamical effects took over to transport the material inward. 
This appears consistent with the long-term observed behaviour of GX~1+4.

\subsection{Mass and angular momentum accretion rates}

The mass and angular momentum rates shown in figures~\ref{fig:mass} 
and~\ref{fig:am} vary relatively smoothly. This is to be expected as
long as the disk is so large that its inner part is a steady viscous
disk, which will smear out any more abrupt changes over a viscous
timescale, which in turn is comparable to the externally imposed
timescale as discussed above. These profiles should be fairly universal
as long as viscosity is important close to the magnetosphere, 
since the non-viscous evolution
simply pushes in a sequence of rings, with the inner one disappearing
by forming a disk and being accreted as in our first simulation. 

The simulations indicate that a reversal of the material accretion
torque should
occur during a period of {\em low} luminosity. There may be some
evidence for this in the data for GX~1+4, especially the non-detection
of the source when its long-time spin rate change reversed in 1984, or
possibly the low luminosities just before and after the 1994-1995
spin-up episode observed by BATSE (Chakrabarty et al. 1997). 
Although the timescales of these two observed features
differ by a factor of 10, it is not inappropriate to
apply the same model to them, since the timescale quoted in
equation~\ref{eq:freefall} 
 for the change of the outer boundary condition is only typical, and
shorter and longer timescales have been seen in calculations of 
the flip-flop model. An alternative explanation put forward by 
Cui (1997) invokes a ``magnetic propeller''
mechanism to explain the low accretion episode of GX~1+4.

We do not intend that this model explain all details of the observations
of torque and luminosity in GX~1+4, which sometimes show the same
torque for very different luminosities even when the disk is
supposedly rotating in the same direction as evidenced by the
short-term torque-luminosity correlation. Clearly an additional model
for the interaction between disk and magnetosphere and the associated
braking torques is required.

\section{Concluding Remarks}

We are of course fully aware of the extreme simplifications that were used
in our simulations. In more realistic studies, 3-dimensional effects
such as subsequent bursts of material arriving at the outer boundary
not being coplanar or not having exactly opposite angular momentum,
or vertical motion due to the strong heating that must
occur in the strong shearing regions should be included. We plan to
address some of these in future work. 

We expect, however, that two main effects will remain standing even
after more detail has been included. The first is that changes at the
magnetospheric radius occur on a similar timescale as 
changes at the outer boundary, and the second is that rings
with opposite rotation do not tend to mix completely but rather create
a gap between themselves, leading to the prediction
that torque reversals should occur during minima in accretion
luminosity.

\acknowledgements
The authors would like to thank Dayal Wickramasinghe and Phil Armitage
for comments and discussions.

\pagebreak
\begin{deluxetable}{lcccccc}
\tablecaption{Parameter values \label{tbl:params}}
\startdata
Parameter & $c_0$ & $h_{\rm max}$ & $r_{\rm in}$ & $r_{\rm circ}$ &
$r_{\rm out}$ & $\Delta t$\nl
Units & \vsc & \lsi & \lsi & \lsi & \lsi & \tsc \nl
Value & 0.1 & 0.02 & 0.05 & 0.5 & 1.2 & 0.10\nl
\enddata
\end{deluxetable}

\end{document}